\documentclass[%
 aip,
 jcp,%
 amsmath,amssymb,
reprint,%
]{revtex4-1}

\usepackage{graphicx}
\usepackage{dcolumn}
\usepackage{bm}
\usepackage[dutch,english]{babel}
\begin{document}


\title{Entangled two photon absorption cross section on the 808 nm region for the common dyes Zinc tetraphenylporphyrin and  Rhodamine B}

\author{Juan. P. Villabona-Monsalve}
 \altaffiliation[Currently at ]{Escuela de Qu\'imica, Universidad Industrial de Santander. Bucaramanga, Santander 680002, Colombia}
\email{juanpablovillabona@hotmail.com}
\author{Omar Calder\'on-Losada}%
\author{Mayerlin Nu\~nez Portela}
\author{Alejandra Valencia}
\affiliation{Laboratorio  de \'Optica  Cu\'antica,  Universidad  de  los  Andes,  A.A.  4976,  Bogot\'a  D.C.,  Colombia}

\date{\today}

\begin{abstract}
We report the measurement of the entangled two photon absorption cross section, $\sigma_E$,  at $808$~nm on  organic chromophores in solution in a low photon flux regime. We performed measurements on Zinc tetraphenylporphyrin (ZnTPP) in Toluene and  Rhodamine B (RhB) in Methanol. This is, to the best of our knowledge, the first time that $\sigma_E$ is measured for RhB. Additionally, we report a systematic study of the dependence of $\sigma_E$ on the molecular concentration for both molecular systems. In contrast to previous experiments, our measurements are based on detecting the pairs of photons that are transmitted by the molecular system. By using a coincidence count circuit it was possible to improve the signal to noise ratio. This type of work is important for the development of spectroscopic and microscopic techniques using entangled photons.
\end{abstract}
\maketitle

\section{\label{sec:level1}INTRODUCTION}

Two photon absorption (TPA) is a non-linear process in which, by absorbing two photons, an specific electronic excited state of a system can be accessed \cite{Rumi:10}.  In this process, energy conservation is satisfied given that the sum of  the energy of the individual photons should be equal to the energy of the transition that is addresed. The TPA phenomena was theoretically proposed by Maria Goeppert Mayer on 1931 \cite{ANDP:ANDP19314010303} and it was experimentally demonstrated after the invention of the laser \cite{PhysRevLett.7.229}. TPA has been studied  using high intensity pulsed lasers on the femtosecond \cite{Xu:96, doi:10.1021/jz400851d, doi:10.1021/ja803268s} and picosecond \cite{Oulianov2001235} regime and it has been extensively used for spectroscopic and microscopic techniques \cite{So2000} as well as in photoinduced phenomena  on a wide variety of materials.\cite{Finikova2007,Chung2001} 

In recent years, an interest to study TPA induced by  sources with different statistical properties than lasers, such as thermal, \cite{Jechow2013} squeezed \cite{PhysRevLett.75.3426} and entangled light,\cite{doi:10.1021/jp066767g, doi:10.1021/ja803268s, doi:10.1021/jz400851d, PhysRevLett.93.023005}  has appeared. Particularly, entangled light has been experimentally tested as a convenient source to induce two-photon transitions on molecules due to its non-classical properties.  One of the motivations for this type of work is the possibility to induce TPA with a low photon flux. This capability may have important implications reducing photodestruction probability and photo-bleaching of the sample, allowing to develop less invasive methods to study TPA in biological samples.\cite{hoover2013,kim2015}  
Entangled light has also been used to induce TPA in semiconductors to study temporal correlations of twin beams \cite{Boitier2011}. From a theoretical perspective, TPA induced by entangled light with different type of frequency correlations has been studied as a tool to learn about the spectral properties of a sample. \cite{Salazar:12, Saleh1998, svozilik2016}

In this paper, we study TPA induced in molecules by entangled light, generated by the process of Spontaneous Parametric Down Conversion (SPDC)\cite{Shih2003} pumped by a cw laser, at a low photon flux.  The probability of entangled TPA in a sample is quantified by means of the entangled TPA cross section, $\sigma_E$. We infer the value of $\sigma_E$ for the commercially available compounds Zinc tetraphenylporphyrin (ZnTPP) and Rhodamine B (RhB) by measuring the absorption signal. Measurements of $\sigma_E$ for ZnTTP have been previously reported \cite{doi:10.1021/jz400851d}; however, this  is, to the best of our knowledge, the first time that $\sigma_E$ is measured for RhB. Additionally, we report, for these two molecules, a systematic study of the dependence of $\sigma_E$ on the molecular concentration. In contrast to previous experiments, \cite{doi:10.1021/jz400851d, doi:10.1021/jp066767g, doi:10.1021/ja803268s} our measurements are based on detecting, by means of a coincidence count circuit, the pairs of photons that are transmitted by the molecular system,  improving the signal to noise ratio.

\section{\label{sec:level1}THEORY}

In this section, we present a theoretical model to obtain $\sigma_E$ from experiments based on coincidence counts.  In general, TPA can be induced by using entangled light and random light sources such as lasers. Considering a molecular system, the TPA rate per molecule, $r_{\text{TPA}}$, can be written as:\cite{PhysRevLett.78.1679, PhysRevB.69.165317, PhysRevA.41.5088}
\begin{equation}\label{rTPA}
r_{\text{TPA}} =  \sigma_{E}\ \phi + \delta_{R}\ \phi^2,
\end{equation}
where $\phi$ is the incident photon flux density (photons s$^{-1}\text{cm}^{-2}$) impinging on the molecule and $\delta_R$ is the random TPA cross section.  The quadratic term in Eq.~(\ref{rTPA}) represents the contribution of random sources to $r_{\text{TPA}}$. Typicaly, $\delta_{R}$ is on the order of $10^2$ GM (1 GM$=10^{-50}$cm$^{4}$~s~photon$^{-1}$~molecule$^{-1}$) and therefore, a high photon flux density ($10^{18}$~photons~s$^{-1}$cm$^{-2}$) is required to induce TPA. The  term $\sigma_{E}\phi$ in Eq.~(\ref{rTPA}) represents the contribution of entangled light to $r_{TPA}$.  A significant difference between random and entangled TPA rates is clearly seen by the dependence of $r_{TPA}$ on the incident photon flux. The linear dependence with $\phi$ for entangled light and the fact that $\sigma_E$ can be on the order of $10^{-18}$~cm$^2$~molecule$^{-1}$, allows to observe TPA with a photon flux density on the order of $10^{12}$~photons~s$^{-1}$~cm$^{-2}$ as reported in Ref. [\onlinecite{doi:10.1021/ja803268s}]. This values for photon flux density are significantly lower than the one for random sources.

Let us consider the case where TPA in a molecule is induce by a  source of entangled light with a rate $r_{\text{ETPA}}$. In this case, the quadratic dependence on the photon flux in Eq.~(\ref{rTPA}) can be neglected and $r_{\text{TPA}}=r_{\text{ETPA}}$.  Considering a source of entangled light that produces pairs of photons, $\phi=2\phi^\prime$, with $\phi^\prime$ denoting the incident entangled photon pair flux density impinging on the molecule. In this case, 
\begin{equation}\label{rETPA}
r_{\text{ETPA}} = 2\sigma_E\phi^\prime.
\end{equation}
 
From the experimental point of view, the quantity $r_{\text{ETPA}}$ can be estimated by measuring the rate of photon pairs absorbed by a molecular system, $R_{abs}$. For a sample containing $N$ molecules on a volume $V$, 
\begin{equation}\label{rETPA2}
r_{\text{ETPA}} =\frac{ 2R_{abs}}{N}.
\end{equation}

Comparing the theoretical expression for $r_{\text{ETPA}}$,  Eq.~(\ref{rETPA}), with the experimental version in Eq.~(\ref{rETPA2}), and taking into account that $N=cVN_{A}$, with $c$ the concentration of molecules in the sample and $N_A$  Avogadro's number,  
\begin{equation}\label{rETPA3}
R_{abs} = cVN_{A}\sigma_E\phi^\prime .
\end{equation}

In particular, in a configuration where the  detection system is based on measuring the coincidence rate of photon pairs, the incident entangled photon pair flux density becomes $\phi^\prime=R/A$, where $R$ is the detected rate of photon pairs produced by the light source that interacts with the molecules in an area $A$. Under these considerations, 
\begin{equation}\label{rETPA4}
R_{abs} = \frac{cVN_{A}\sigma_E}{A}R .
\end{equation}

For molecules studied in liquid solution, it is important to account for the signal scattered by the solvent. In order to do this, it is necessary to measure the rate of photon pairs transmitted through the solvent, $R_{solvent}$, that becomes the incident entangled photon pair flux impinging on the molecules ($R\rightarrow R_{solvent}$); therefore, Eq.~(\ref{rETPA4}) becomes
\begin{equation}\label{rETPA5}
R_{abs} = \frac{cVN_{A}\sigma_E}{A}R_{solvent} .
\end{equation}

Equation~(\ref{rETPA5}) allows to estimate $\sigma_{E}$ for a molecule from variables that can be experimentally controlled. By measuring $R_{abs}$ as a function of $R_{solvent}$ and performing a linear fit, it is possible to estimate $\sigma_E$ for a given molecule at a particular concentration.  The experimental procedure to infer $\sigma_E$ will be described in detail in the following sections.

\section{\label{sec:level1}EXPERIMENT}

\begin{figure*}[htb]
\centering
\includegraphics[width=0.85\textwidth]{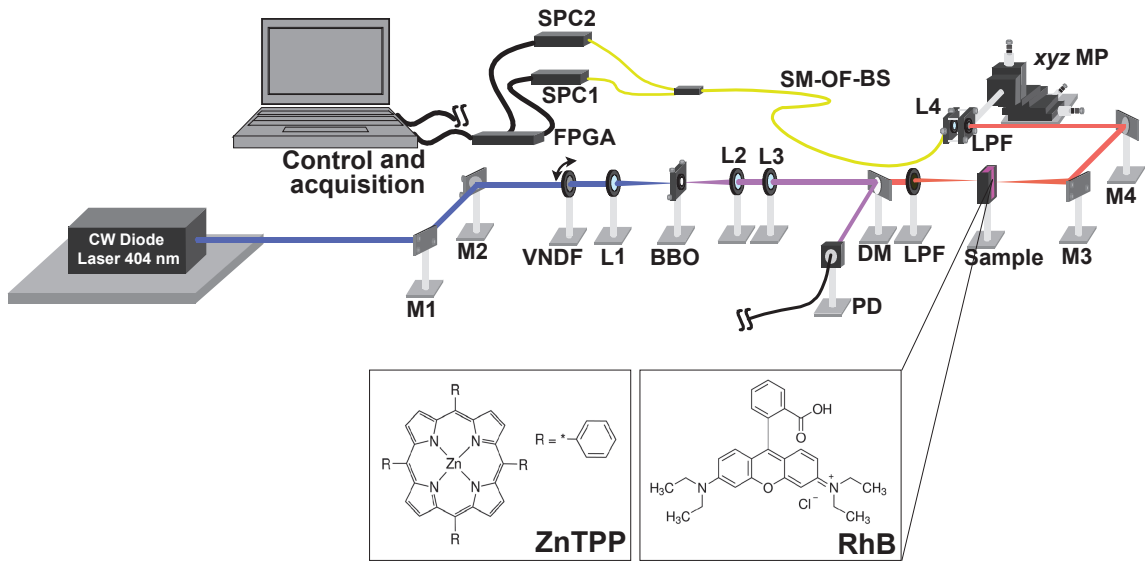}
\caption{Experimental setup to measure entangled TPA for different molecules.  The entangled light source is based on SPDC produced at a BBO crystal pumped by a cw laser. Different lenses, L, are used to control the spatial shape of the light. Mirrors, M, are used to guide the light in the experiment. SPDC photons are couple to a fiber beam splitter, SM-OF-BS, and then detected by single photon counters, SPC1 and SPC2. The molecules at the bottom are the ones studied in this work.} \label{setup}
\end{figure*}

In this section, we describe  the experiment to obtain $\sigma_E$ for two different molecular systems. The experimental setup is shown in Fig.~\ref{setup}. Entangled photon pairs are produced by SPDC, send into a sample that can be ZnTPP or RhB, and finally detected by single photon counters that are connected to a FPGA (Field Programmable Gate Array)  to measure single and coincidence count rates.  In the following, a detailed description of the entangled light source and the molecular systems will be presented.

\subsection{\label{sec:source}Entangled light source}

The entangled light to induce TPA in molecules was generated by the process of SPDC on a non-linear crystal.  Roughly speaking, SPDC is a process in which a photon from a pump beam with a wavelength $\lambda_p$ impinges on a non-linear crystal and occasionally, pairs of photons, known as signal and idler, are produced with wavelengths $\lambda_s$ and $\lambda_i$, respectively. In our setup, Fig.~\ref{setup}, a cw laser at 404 nm pumps a 1 mm BBO ($\beta$-Barium Borate) type I crystal to produce SPDC photons on a degenerate and collinear configuration center at $\lambda_{s}^0=\lambda_{i}^0=808$ nm. The pump is focused into the crystal by L1 ($f_1=50$ mm). The pump power is monitored by a photodiode (PD) and it is tuned from  1 mW to 20 mW by means of a motorized variable neutral density filter (VNDF)  Thorlabs NDC-50C-4M. After the BBO, the SPDC light is collected by L2 ($f_2=50$ mm) and then focused by L3 ($f_3=300$ mm) to produce a beam waist $w_{0}=61~\mu$m on the position of the sample. Following L3,  a dichroic mirror (DM) and  a long pass filter (LPF), with cutoff wavelength at 750 nm, are used to remove the residual pump beam. After passing through the sample, the SPDC photons are coupled by using L4 ($f_4=11$ mm) into a single mode optical fiber-beamsplitter (SM-OF-BS) from OZ-optics. L4 is mounted on a three axis micrometric positioner ($xyz$ MP) that has an extra LPF to reduce the detection of the residual pump beam. The photons from the SM-OF-BS outputs are detected by SPC1 and SPC2, single-photon counters (Perkin-Elmer, SPCM-AQR-13). The output from the detectors is electronically analyzed by the FPGA to obtain the rate of single and coincidence counts in a 9 ns time window. The single photon count rate from our source at the maximum pump power is on the order of $5\times10^5$~photons~s$^{-1}$, leading to a entangled photon flux density on the order of $10^{11}$~photons~cm$^{-2}$~s$^{-1}$.

\subsection{\label{sec:sample} Molecular systems}

The measurement of $\sigma_E$ was performed on the commercially available compounds Zinc tetraphenylporphyrin (ZnTPP) and Rhodamine B (RhB). ZnTPP was obtained from meso-Tetraphenylporphhyrin ($\geq$95 \% purity, Sigma-Aldrich) by the method reported on Ref.  [\onlinecite{Metalloporphyrins}] and solved on high performance liquid chromatography (HPLC) grade Toluene. RhB ($\geq$95 \% purity, Sigma-Aldrich) was used as received and solved on HPLC grade Methanol. The sample was placed in our setup as indicated in Fig.~\ref{setup} using a quartz cuvette of 10 mm pathlength. Taking into account that the waist of the SPDC photons in the sample leads to a Rayleigh range of $14$~mm, the interaction volume can be considered as a cylinder with transverse area and volume on the order of $2\times10^{-4}$~cm$^{2}$ and  $2\times10^{-4}$~cm$^{3}$, respectively. We prepared different concentrations for the molecules in the corresponding solvent. The concentration values were obtained by measuring the absorption spectra using a Specord 50 Plus spectrophotometer (Analytik Jena) and the reported values for the extinction coefficients.\cite{P19750001401, Nag2009188}

In previous reports, the value of $\sigma_E$ and $\delta_r$ have been determined for ZnTPP.\cite{doi:10.1021/jz400851d}  In these measurements entangled light was produced by SPDC generated by a pulsed laser and, before interacting with the sample, it was filtered using a 20 nm bandpass filter around 800 nm. The detection system in their case was based only on single counts. On the other hand, values of $\delta_r$ for RhB have been reported for wavelengths around 800 nm;\cite{Makarov:08,Nag2009188} however, the value of $\sigma_E$ has never been reported. In our experimental setup, the molecular system was illuminated with the whole spectrum of the SPDC photons centered at 808 nm with a bandwidth of 130 nm full width half maximum since we did not use band pass filters. 

Besides their importance for spectroscopic and microscopic techniques, the selected molecules for this study have shown relevance for different applications such as optical sensors.\cite{doi:10.1021/ja803268s} This versatility is based on their photophysics, characterized by high fluorescence quantum yields and long lifetimes for the first electronic excited state.\cite{doi:10.1021/jp0203999, Lukaszewicz2007359, Nag2009188, Kristoffersen2014}

\section{\label{sec:level1}RESULTS AND DISCUSSION}
\begin{figure}[h!] 
\centering
\includegraphics[width=0.5\textwidth]{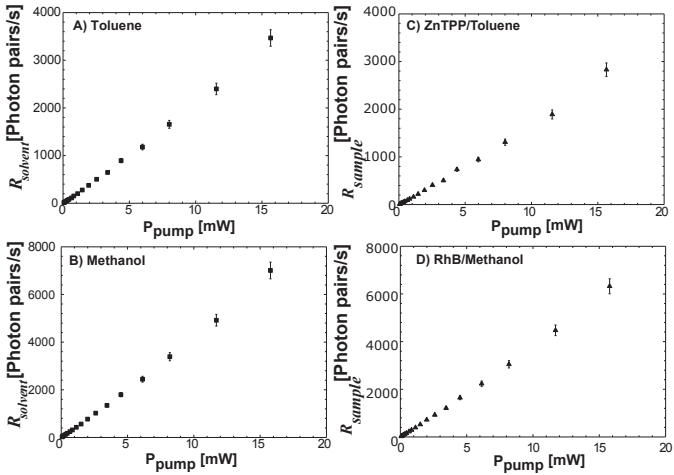}
\caption{ Experimental data for $R_{solvent}$ and $R_{sample}$ as a function of $P_{pump}$. Panel A corresponds to Toluene and panel B corresponds to Methanol. Panel C shows the behavior of $R_{sample}$ for ZnTPP in Toluene with a concentration of $c=63~\mu$M and panel D shows this behavior for RhB in Methanol with a concentration of $c=4.5$ mM.}\label{pump} 
\end{figure}
In order to get a value of $\sigma_E$,  it is necessary to measure the rate of photon pairs absorbed by a molecular system, $R_{abs}$, as a function of $R_{solvent}$, as stated in Eq.~(\ref{rETPA5}). For molecules in a solvent, $R_{abs}=(R_{solvent}-R_{sample})$, where $R_{sample}$ is  the detected rate of photon pairs that passes through the sample, i.e.,  through solvent plus molecular system.  For a fixed concentration, volume and area, it is possible to measure $R_{solvent}$ and $R_{sample}$ as a function of  the power of the pump beam, $P_{pump}$, that generates the entangled photon pairs. These results are presented in Fig.~\ref{pump}A and Fig.~\ref{pump}B for Toluene and Methanol and in Fig.~\ref{pump}C and Fig.~\ref{pump}D for ZnTPP and RhB for a given concentration in their corresponding solvents. Each point on Fig. \ref{pump} corresponds to an average of 60 measurements of one second each.  Error bars represent the standard deviation for each measurement. A linear dependence is observed in these graphs. In particular,  Fig.~\ref{pump}A and Fig.~\ref{pump}B allow to find the correspondence between $P_{pump}$ and $R_{solvent}$. 

\begin{figure}[h!] 
\centering
\includegraphics[width=0.45\textwidth]{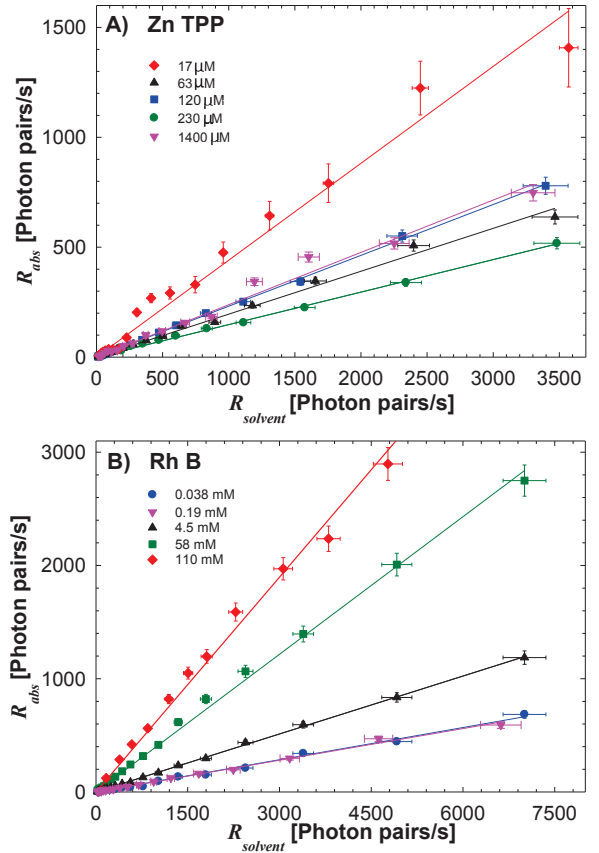}
\caption{Experimental data for $R_{abs}$ as a function of $R_{solvent}$ for different concentrations of the studied molecules. Panel A corresponds to ZnTPP in Toluene and panel B corresponds to RhB in Methanol. The solid lines are linear fits to the data. The fitting parameters are used to calculate $\sigma_{e}$ for each concentration.}\label{results} 
\end{figure}
For a particular sample, $R_{abs}$ as a function of $R_{solvent}$  is obtained by using the correspondence between $P_{pump}$ and $R_{solvent}$ and subtracting $R_{sample}$ from $R_{solvent}$. Fig.~\ref{results} shows the curves $R_{abs}$ as a function of $R_{solvent}$ for different concentrations of the studied molecules. Concentration values for both ZnTPP and RhB samples were chosen to have a $R_{abs}$ signal bigger than 10\% of $R_{solvent}$. Fig.~\ref{results}A depicts the results for ZnTPP and Fig.~\ref{results}B for RhB. The data reported corresponds only to a TPA signal. For the low concentrations in the samples, scattering effects are mainly due to the solvent. By subtracting $R_{sample}$ from $R_{solvent}$, the effects of the scattering are removed.  Absorption signals on the order of $1000$ photon pairs~s$^{-1}$ can be observed for different concentrations at different values of $R_{solvent}$. At low values of $R_{solvent}$ (50 photon pairs s$^{-1}$), the measured values for $R_{abs}$ are less than 20 photon pairs s$^{-1}$, indicating a good signal to noise ratio. 

The solid lines in Fig.~\ref{results} correspond to linear fits to the data. The fits were force to pass through the origin since the background noise was minimized due to the characteristics of the entangled light source and the coincidence detection system. 
 According to Eq.~(\ref{rETPA5}), the slope of the fit is related to the product $\sigma_Ec$. Therefore, there are different values of $\sigma_E$ and $c$ that can turn into the same slope. This is the case for the concentrations $120~\mu$M and $1400~\mu$M of ZnTPP in Fig. \ref{results}A and for the concentrations of $0.038$ mM and $0.19$ mM of RhB in Fig. \ref{results}B. 

\begin{figure}[h!] 
\centering
\includegraphics[width=0.40\textwidth]{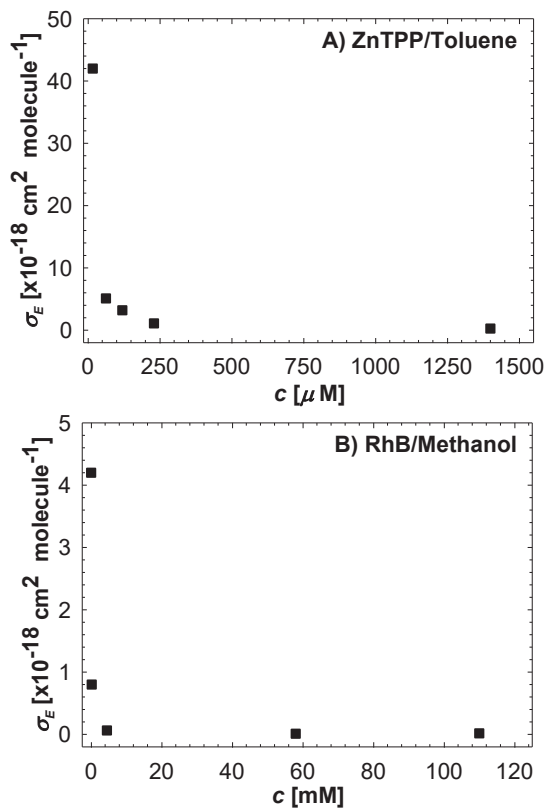}
\caption{Dependence of $\sigma_E$ with the concentration of molecules in a solvent. Panel A corresponds to ZnTPP in Toluene and panel B corresponds to RhB in Methanol.}\label{concentrationplot}
\end{figure}

The results of $\sigma_E$ for different concentrations are summarized in Table~\ref{table1} and plotted in Fig.~\ref{concentrationplot}A for ZnTPP and in Fig.~\ref{concentrationplot}B for RhB. A strong decay is observed for  $\sigma_E$ when the concentration increases in Fig.~\ref{concentrationplot}. This behavior can be understood by an analogy with previous reports on the dependence of $\delta_r$ and $\sigma_E$ with the concentration. \cite{Ajami2015524,doi:10.1021/ja803268s} The decreasing of $\sigma_E$ for high concentrations can be interpreted by considering aggregation of the molecules that lead to screening effects. 

\begin{table}[htb] 
\centering
\begin{tabular}{|c | c | }
\multicolumn{2}{c}{ZnTPP}\\
\hline
$c$ ($\mu M$) & $\sigma_{E}\times 10^{-18}$ (cm$^{2}$~molecule$^{-1}$)\\
\hline
$17$ & $42\pm~5.2$\\
$63$ & $5.1\pm~ 0.46$\\
$120$ & $3.2\pm~ 0.20$\\
$230$ & $1.1\pm~ 0.07$\\
$1400$ & $0.27\pm~ 0.026$\\
\hline 
\multicolumn{2}{c}{RhB} \\
\hline
$c$ (mM) & $\sigma_{E}\times 10^{-18}$ (cm$^{2}$~molecule$^{-1}$)\\
\hline
$0.038$ & $4.2\pm~ 0.34$\\
$0.19$ & $0.80\pm~ 0.068$\\
$4.5$ & $0.063\pm~ 0.0039$\\
$58$ & $0.011\pm~ 0.00084$\\
$110$ & $0.017\pm~ 0.0018$\\
\hline
\end{tabular}
\caption{Values of  $\sigma_{E}$ for different concentrations obtained from the fitting parameters in Fig.~\ref{results}. }\label{table1}
\end{table}

\section{\label{sec:level1}CONCLUSIONS}

We reported on the measurement of the entangled two photon absorption cross section, $\sigma_E$, for ZnTPP and RhB. The value of $\sigma_E$ for ZnTPP agrees with previously reported results. For RhB, this is the first time that a measurement for $\sigma_E$ has been performed. Additionally, we showed that sample concentration has important effects on $\sigma_E$. The data we reported was taken in an experimental setup with two main features that are different from previous setups to measure $\sigma_E$. First, our entangled light source was based on SPDC  pumped by a cw laser instead of a pulsed laser. Second, the detection system we used was based on counting the rate of photon pairs absorbed by the molecules instead of the number of single photons absorbed. This last feature, allowed us to show that regardless the low photon flux density of entangled photons ($10^{11}$~photons~cm$^{-2}$~s$^{-1}$), the TPA processes can be induced and the number of absorbed photon pairs can be measured. The obtained experimental results provide support for the implementation of spectroscopic and microscopic techniques based on the process of entangled two-photon absorption on molecules. Additionally, this type of work is beneficial for studying biological samples and the implementation of detection systems operating at low photon fluxes.

\section*{ACKNOWLEDGMENTS}
This work was financially supported by Facultad de Ciencias, Universidad de los Andes, Bogot\'a, Colombia. The authors thank Dr. Yenny Hern\'andez at Universidad de los Andes for fruitful discussions and Dr. Gilma Granados at Universidad Nacional de Colombia for the synthesis and purification of Zinc tetraphenylporphyrin.

\end{document}